\documentclass[a4wide]{article}
\usepackage{RRA4}
\RRNo{6470}
\usepackage{hyperref}
\usepackage{lscape}
\usepackage{rotating}
\RRdate{Fevrier 2008}

\RRauthor{
	Pedro T. Monteiro
	\thanks[ucb]{Universit\'{e} Claude Bernard Lyon 1, 43 Bvd. du 11 Nov. 1918, 69622 Villeurbanne Cedex, France}%
	\thanks[inria]{INRIA Grenoble - Rh\^one-Alpes, 655 Av. de l'Europe, Montbonnot, 38334 St. Ismier Cedex, France}%
	\thanks[inesc]{IST/INESC-ID, 9 Rua Alves Redol, 1000-029 Lisbon, Portugal}%
\and
	Delphine Ropers\thanksref{inria}
\and
	Radu Mateescu\thanksref{inria}%
\and
	Ana T. Freitas\thanksref{inesc}%
\and
	Hidde de~Jong\thanksref{inria}%
}
\authorhead{Monteiro \& Ropers \& Mateescu \& Freitas \& de~Jong}
\RRtitle{Patterns en Logique Temporelle pour faire des requ\^etes \`a mod\`eles dynamiques de r\'eseaux d'interaction cellulaires}
\RRetitle{Temporal Logic Patterns for Querying Dynamic Models of Cellular Interaction Networks}
\titlehead{Patterns for Querying Models of Cellular Interaction Networks}
\RRresume{
Les mod\`eles de la dynamique des r\'eseaux d'interaction cellulaires sont devenus de plus en plus grands au cours des derni\`eres ann\'ees. La v\'erification formelle bas\'ee sur le model checking fournit une technologie puissante pour faire face \`a cette augmentation en taille et en complexit\'e. Malheureusement l'application de telles approches est limit\'ee par la difficult\'e pour l'utilisateur non exp\'eriment\'e de formuler des questions appropri\'ees en logique temporelle.
Pour faire face \`a ce probl\`eme, nous proposons l'utilisation de patterns, des squelettes de requ\^etes \`a haut niveau qui expriment des questions biologiques r\'ecurrentes et qui peuvent \^etre automatiquement traduites en logique temporelle.
L'applicabilit\'e de l'ensemble de patterns d\'evelopp\'es a \'et\'e \'etudi\'ee par l'analyse d'une extension du mod\`ele du r\'eseau des r\'egulateurs qui contr\^ole la r\'eponse au manque de carbone chez \textit{Escherichia coli}.
}
\RRabstract{
Models of the dynamics of cellular interaction networks have become increasingly larger in recent years. Formal verification based on model checking provides a powerful technology to keep up with this increase in scale and complexity. The application of model-checking approaches is hampered, however, by the difficulty for non-expert users to formulate appropriate questions in temporal logic.
In order to deal with this problem, we propose the use of patterns, that is, high-level query templates that capture recurring biological questions and that can be automatically translated into temporal logic. The applicability of the developed set of patterns has been investigated by the analysis of an extended model of the network of global regulators controlling the carbon starvation response in \textit{Escherichia coli}.
}
\RRmotcle{V\'erification formelle, R\'eseaux de r\'egulation g\'enique, Biologie des syst\`emes, Simulation qualitative, Logique temporelle}

\RRkeyword{Formal verification, Genetic regulatory networks, Systems biology,   Qualitative Simulation, Temporal logic}
\RRprojets{HELIX et VASY}
\RRtheme{\THBio}
 \URRhoneAlpes 
 \RCGrenoble 

\newtheorem{df}{Definition}
\usepackage{amsmath,amsfonts}
\begin{document}
\makeRR   

\section{Introduction}

Models of the dynamics of cellular interaction networks have become increasingly larger in recent years. While whole-cell models are not on the horizon yet, complex networks underlying specific cellular processes have been modeled in detail, such as the osmotic shock response in yeast \cite{HdJ2836}, the yeast cell cycle \cite{HdJ2457}, and signalling pathways involved in cancer \cite{HdJ2590}. The study of these models by means of analysis and simulation tools leads to large amounts of predictions, typically time-courses of the concentrations of several dozens of molecular components in a variety of physiological conditions and genetic backgrounds. This raises the question how to make sense of these data, that is, how to obtain an understanding of the way in which particular molecular mechanisms control the cellular process under study, and how to identify interesting predictions of novel phenomena that can be confronted with experimental data.

Methods from the field of formal verification provide a  promising way to deal with the analysis of large and complex models of cellular interaction networks \cite{HdJFisher}. Generally speaking, formal verification proceeds by specifying dynamical properties of interest as statements in temporal logic. Efficient so-called model-checking algorithms, implemented in publicly-available computer tools, exist to determine whether the statements are true or false, and thus whether the dynamic properties are satisfied by the model. The methods are generally applicable to discrete models of cellular interaction networks, or continuous models that have been discretized under a suitable abstraction criterion. Several examples exist of the application of model-checking approaches in systems biology (\textit{e.g.}, \cite{Antoniotti,HdJ2427,HdJ2379,HdJCalder,HdJ2380,Fisher}).

Formal verification based on model checking provides a powerful technology to query models of cellular interaction networks. It raises a number of new issues though, notably that of formulating good questions when analyzing a huge network model. The problem of posing relevant and interesting questions is critical in modeling in general, but even more so in the context of applying formal verification techniques, due to the fact that it is not easy for non-experts to formulate queries in temporal logic.
For instance, the property ``Gene \textit{g} is eventually expressed, and necessarily preceded over the whole duration of the experiment by a concentration larger than 0.9 $\mu$M of the transcription factor P'' corresponds to the following CTL formula, where $exp_g$ denotes expression of \textit{g}:

\begin{equation}
EF (exp_g) \wedge \neg E (True ~U~ (\neg ([P]>0.9 \mu M) \wedge E (True ~U~ exp_g))) 
\end{equation}

The response to this problem proposed by the formal verification community is the use of patterns, that is, high-level query templates that capture recurring questions in a specific application domain and that can be automatically translated to temporal logic \cite{HdJ2605}. Apart from lists of example queries \cite{HdJ2380}, the systematic definition of queries has not received any attention in systems biology thus far. 

The aim of this paper is to develop a set of patterns for the analysis of dynamic models of cellular interaction networks. Its main contributions lie, first, in the development of generic query templates, based on a review of frequently-asked questions by modelers, and the translation of these templates into temporal logic formulas (Sec.~\ref{sec:patterns}). Second, we apply the patterns for analyzing the qualitative dynamics of a large and complex model of the \textit{E. coli} carbon starvation response (Sec.~\ref{sec:bioapplic}).
This model extends a previous model \cite{BioSystems} by taking into account additional regulators of the bacterium, notably a module centered around the general stress response factor RpoS. We verify the control the latter is predicted to exert on the DNA supercoiling level in the cell.

\section{Patterns of biological queries}
\label{sec:patterns}

\subsection{Description of network dynamics}

As a basic hypothesis, we assume that the dynamics of molecular
interaction networks can be modeled by means of {\em finite state
transition systems} (FSTSs) \cite{Clarke99}. The latter formalism provides a general description of a dynamical system that implicitly
or explicitly underlies many of the existing discrete formalisms used
to model cellular interaction networks, such as Boolean networks and
their generalizations, Petri nets, and process algebras. In addition,
by defining appropriate discrete abstractions, continuous models of
cellular interaction networks can also be mapped to FSTSs. The
generality of the FSTS formalism is important for assuring the wide
applicability of the patterns developed in this section. Moreover,
statements in temporal logics are usually interpreted on FSTSs, so
that the latter naturally connect network models to model-checking
tools.

A finite state transition system is formally defined as a tuple
$\Sigma = \langle S, AP, L, T, S_0 \rangle$, where $S$ is a set
of states, $AP$ is a set of atomic propositions, $L: S \rightarrow
2^{AP}$ is a labeling function that associates to a state $s\in S$ the
set of atomic propositions satisfied by $s$, $T \subseteq S \times S$
is a relation defining transitions between states, and $S_0\subseteq
S$ is a set of initial states. For our purpose, $S$ describes the
possible states of the cellular interaction network, each of which is
characterized by a set of atomic propositions, such as that the
concentration of protein P is increasing, or that the concentration of
metabolite M is smaller than 5~mM.

\subsection{Identification of patterns}

The notion of patterns originates in architecture~\cite{Alexander77}
and was introduced in the domain of software engineering as a means to
capture expert solutions to recurring problems in program design \cite{Gamma}.
In the formal verification domain they have been introduced in an influential paper
by \cite{HdJ2605}, to help non-expert users formulate their
temporal-logic queries. In the latter context, patterns are high-level
descriptions  of frequently asked questions in an
application domain that are formulated in structured natural language
rather than temporal logic.
The aim of the patterns is not to cover all possible
questions an expert can think of, but rather to simplify the formulation of those that are primary.

The difficulty of proposing patterns is to come up with a limited
number of query schemas that are sufficiently generic to be applicable
in a variety of situations, and at the same time sufficiently
concrete to be comprehensible for the non-expert user. Moreover, the
overlap between the patterns should be minimal. We analyzed a large
number of modeling studies in systems biology (starting from the references in \cite{HdJ2482}), as well as previous
applications of model checking and temporal logic (\textit{e.g.},
\cite{Antoniotti,HdJ2427,HdJ2379,HdJCalder,HdJ2380,Fisher}). This bibliographic research allowed us to identify an open-ended list of questions on the dynamics of genetic, metabolic, and signal transduction networks, for instance:

\begin{itemize}
\item Is the basal glycerol production level combined with rapid closure of Fps1 sufficient to explain an initial glycerol accumulation after osmotic shock? \cite{HdJ2836}
\item Once a cell has executed START, does it slip back into G1 phase and repeat START? Or rather, must it execute a FINISH to return to G1? \cite{HdJ2457}.
\item Does Shc phosphorylation exhibit a relative acceleration with decreasing EGF concentration and show a decline over time? \cite{HdJ2590}
\end{itemize}

The identified questions were grouped into
four categories, depending on whether they concerned the {\em occurrence}/{\em exclusion}, {\em consequence}, {\em sequence}, and {\em invariance} of cellular events. For each of these, we developed an appropriate pattern, capturing the essence of the question and the most relevant variants.

\subsection{Description of patterns}
\label{sec:patterns,descr}

The patterns consist of structured natural language phrases,
represented in schematic form, with placeholders for so-called {\em state
descriptors}. A state descriptor is a statement expressing a state property, and takes the form of (a Boolean
combination of) atomic propositions. Let $\phi,\psi$ be state
descriptors, then

\begin{align}
\phi, \psi &::= p_1 \in AP | p_2 \in AP | \ldots \nonumber \\
&::= \neg \phi | \phi \wedge \psi | \phi \Rightarrow \psi| \ldots \nonumber
\end{align}

The state descriptors are interpreted on the FSTS, in the sense that
their meaning is formally defined as the set of states $S_1 \subseteq
S$ satisfying the state descriptor. In addition to (Boolean combinations of) atomic propositions,
the state descriptors may be temporal-logic formulas defined
on the atomic propositions. We will return to this generalization in Sec.~\ref{sec:discussion}.

It is often convenient to introduce predefined state descriptors that
capture Boolean combinations of atomic propositions that are
recurrently used. Some examples of predefined state descriptors that
we found useful are the following:

\begin{itemize}
\item $Increases_i$/$Decreases_i$: the concentration of molecular component $i$ increases/decreases in this state;
\item $IsSteadyState$: the concentrations of all molecular components are steady in this state;
\item $IsOscillatoryState$: the concentrations of some molecular components oscillate in this state.
\end{itemize}

Notice that the precise definition of the state descriptors depends on
the particular type of FSTS that is used, as the latter determines the
set of atomic propositions $AP$.

\begin{df}[Occurrence/exclusion pattern]\rm ~\\ ~\\
\label{pattern:occurrence}
\centerline{\scalebox{0.75}{\begin{picture}(0,0)%
\includegraphics{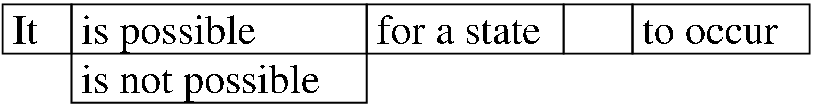}%
\end{picture}%
\setlength{\unitlength}{4144sp}%
\begingroup\makeatletter\ifx\SetFigFontNFSS\undefined%
\gdef\SetFigFontNFSS#1#2#3#4#5{%
  \reset@font\fontsize{#1}{#2pt}%
  \fontfamily{#3}\fontseries{#4}\fontshape{#5}%
  \selectfont}%
\fi\endgroup%
\begin{picture}(3714,486)(1744,-4135)
\put(4411,-3841){\makebox(0,0)[lb]{\smash{{\SetFigFontNFSS{12}{14.4}{\rmdefault}{\bfdefault}{\updefault}$\phi$}}}}
\end{picture}%
}}
\end{df}

This pattern represents the concepts of {\em occurrence} and its
negation, {\em exclusion}. It will often be used during the
development of a model to check for the presence or absence of some
property that was experimentally observed. 
For instance, ``It is possible for a
state with high expression of protein P$_1$ to occur''.
Using this pattern, we can also check for {\em mutual exclusion},
by using the pattern negative form in combination with a
conjunctive state descriptor. For instance, ``It is not possible for a
state to occur in which protein P$_1$ and protein P$_2$ are highly expressed''.

More generally, the {\em exclusion} pattern captures the 
{\em safety properties} used in the domain of concurrent systems.
A {\em safety} property (of which {\em mutual exclusion} is a typical example)
expresses that ``something bad never happens'' during the
execution of the system, in our example
a bad state violating mutual exclusion.

\begin{df}[Consequence pattern]\rm ~\\ ~\\
\label{pattern:consequence}
\centerline{\scalebox{0.75}{\begin{picture}(0,0)%
\includegraphics{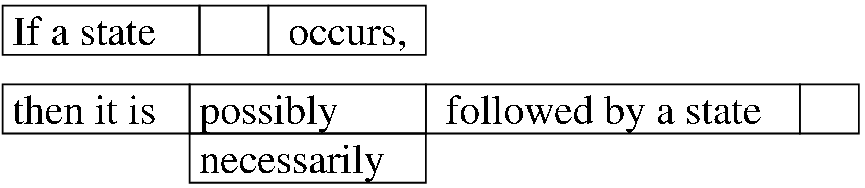}%
\end{picture}%
\setlength{\unitlength}{4144sp}%
\begingroup\makeatletter\ifx\SetFigFontNFSS\undefined%
\gdef\SetFigFontNFSS#1#2#3#4#5{%
  \reset@font\fontsize{#1}{#2pt}%
  \fontfamily{#3}\fontseries{#4}\fontshape{#5}%
  \selectfont}%
\fi\endgroup%
\begin{picture}(3939,846)(844,-1390)
\put(1846,-736){\makebox(0,0)[lb]{\smash{{\SetFigFontNFSS{12}{14.4}{\rmdefault}{\bfdefault}{\updefault}$\phi$}}}}
\put(4546,-1096){\makebox(0,0)[lb]{\smash{{\SetFigFontNFSS{12}{14.4}{\rmdefault}{\bfdefault}{\updefault}$\psi$}}}}
\end{picture}%
}}
\end{df}

The {\em consequence} pattern relates two events separated in
time. More precisely, it expresses that if the first state occurs,
then it is possibly or necessarily followed by the occurrence of the
second state. If the latter state necessarily follows, then the
consequence pattern expresses a form of causal relation. Instances of
this pattern are, for instance, ``If a state occurs in which protein P
is phosphorylated, then it is possibly followed by a state in which
the expression of gene {\em g} decreases'', or ``If a state occurs in
which the concentration of protein P is below 5~$\mu$M, then it is
necessarily followed by a state in which the expression of gene
{\em g} is at its basal level''.

\begin{df}[Sequence pattern]\rm ~\\ ~\\
\label{pattern:sequence}
\centerline{\scalebox{0.75}{\begin{picture}(0,0)%
\includegraphics{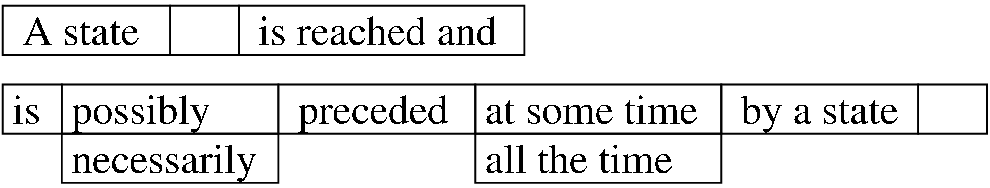}%
\end{picture}%
\setlength{\unitlength}{4144sp}%
\begingroup\makeatletter\ifx\SetFigFontNFSS\undefined%
\gdef\SetFigFontNFSS#1#2#3#4#5{%
  \reset@font\fontsize{#1}{#2pt}%
  \fontfamily{#3}\fontseries{#4}\fontshape{#5}%
  \selectfont}%
\fi\endgroup%
\begin{picture}(4524,846)(1294,-1570)
\put(2161,-916){\makebox(0,0)[lb]{\smash{{\SetFigFontNFSS{12}{14.4}{\rmdefault}{\bfdefault}{\updefault}$\psi$}}}}
\put(5536,-1276){\makebox(0,0)[lb]{\smash{{\SetFigFontNFSS{12}{14.4}{\rmdefault}{\bfdefault}{\updefault}$\phi$}}}}
\end{picture}%
}}
\end{df}

The {\em sequence} pattern represents an ordering relation between
two events. It ought not to be confused with the {\em consequence}
pattern, since the conditional occurrence of the second state
which characterizes the latter, is absent in the {\em sequence}
pattern. Both the first and the second state, in that order, have to
be observed for an instance of the {\em sequence} pattern to be true. 

Four variants of the pattern are distinguished, depending on whether the
second state follows possibly or necessarily after the first state,
and whether the system is in the first state all the time or only
at some time before the occurrence of the second state. Instances of this pattern
are ``A state in which reactions R$_1$ and R$_2$ occur at a high rate
is reached after 2 min, and is possibly preceded at some time by a
state in which the transcription factor P is phosphorylated'' or ``A
steady state is reached and is necessarily preceded all the
time by a state in which nutrient N is absent''.

\begin{df}[Invariance pattern]\rm ~\\ ~\\
\label{pattern:invariance}
\centerline{\scalebox{0.75}{\begin{picture}(0,0)%
\includegraphics{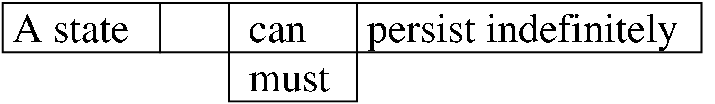}%
\end{picture}%
\setlength{\unitlength}{4144sp}%
\begingroup\makeatletter\ifx\SetFigFontNFSS\undefined%
\gdef\SetFigFontNFSS#1#2#3#4#5{%
  \reset@font\fontsize{#1}{#2pt}%
  \fontfamily{#3}\fontseries{#4}\fontshape{#5}%
  \selectfont}%
\fi\endgroup%
\begin{picture}(3261,474)(3994,-3538)
\put(4816,-3256){\makebox(0,0)[lb]{\smash{{\SetFigFontNFSS{12}{14.4}{\rmdefault}{\bfdefault}{\updefault}$\phi$}}}}
\end{picture}%
}}
\end{df}

The {\em invariance} pattern is used to check if the system can or
must remain indefinitely in a state.
In contrast with the 
{\em occurrence/exclusion} pattern, the question is not whether a particular 
state can be reached, but rather whether a particular state is invariable.
Instances of the pattern are ``A state in which reaction R
occurs at a high rate can persist indefinitely'' and ``A state with a
basal expression of gene {\em g} must persist indefinitely''.

\subsection{Translation to temporal logic}
\label{sec:translation2tl}

By defining a translation into temporal logic of the patterns, the
user queries can be automatically cast in a form that allows the
verification of the specified property by means of model-checking
tools. The patterns defined above are independent of a particular
temporal logic, which allows the same high-level specification of a user query to be verified by means of different approaches and tools. It is worth noticing though that some of the patterns we propose have a branching-time nature (\textit{e.g.}, the {\em consequence} and the {\em sequence} patterns), and therefore these are not translatable into a linear-time formalism, such as LTL \cite{Clarke99}.

Two examples of translations of the patterns in
Sec.~\ref{sec:patterns,descr} are shown in tabular form: the Computational Tree Logic (CTL)
translation and the $\mu$-calculus
translation (Table~\ref{translation:ctl:mu}). In both CTL and $\mu$-calculus, formulas are built upon
atomic propositions. Also, the usual connectors of propositional logic, such as negation ($\neg$),
logical or ($\vee$), logical and ($\wedge$) and implication ($\Rightarrow$),
can be used in both logics.
In addition, CTL provides two types of operators: \textit{path quantifiers},
\textbf{E} and \textbf{A}, and \textit{temporal operators}, such as \textbf{F}
and \textbf{G}.
Path quantifiers are used to specify that a property $p$ is
satisfied by some (\textbf{E}~$p$) or every (\textbf{A}~$p$) path starting
from a given state.
Temporal operators are used to specify that, given a state and a path starting
from that state, a property $p$ holds for some (\textbf{F}~$p$) or for every
(\textbf{G}~$p$) state of the path.
Each path quantifier must be paired with a temporal operator.
In the case of $\mu$-calculus, two types of operators are provided: {\em fixed points}, the least ($\mu$) and greatest ($\nu$), and {\em modal operators}, possibility ($\lozenge$) and necessity ($\square$).
Least and greatest fixed points specify finite and infinite recursive applications of a formula, respectively. For instance, given a state and a path starting from that state, the fact that a property $p$ holds for some state or for all states of the path is expressed using a least ($\mu$) or a greatest ($\nu$) fixed point, respectively.
Modal operators are used to specify that, given a state, a property $p$ possibly ($\lozenge~p$) or necessarily ($\square~p$) holds on some or all of its outgoing states.

\begin{landscape}
\begin{table}
{\small
\begin{tabular}{|l|l|l|}
\hline
\textbf{Occurrence/Exclusion pattern} & \textbf{CTL} & \textbf{$\mu$-calculus} \\ \hline
It \textbf{is possible} for a state $\phi$ to occur & $EF~ (\phi)$ & $\mu X.(\phi \vee \lozenge X)$ \\ 
It \textbf{is not possible} for a state $\phi$ to occur & $\neg EF~ (\phi)$ & $\neg \mu X.(\phi \vee \lozenge X)$ \\ \hline 
\hline

\multicolumn{3}{|l|}{\textbf{Consequence pattern}} \\ \hline
If a state $\phi$ occurs, then it is \textbf{possibly} followed by a state $\psi$ & $AG~ (\phi \Rightarrow EF ~(\psi))$ & $\nu X.((\phi \Rightarrow \mu Y . (\psi \vee \lozenge Y)) \wedge \square X)$ \\
If a state $\phi$ occurs, then it is \textbf{necessarily} followed by a state $\psi$ & $AG~ (\phi \Rightarrow AF ~(\psi))$ & $\nu X . ((\phi \Rightarrow \mu Y . (\psi \vee \square Y)) \wedge \square X)$ \\ \hline
\hline

\multicolumn{3}{|l|}{\textbf{Sequence pattern}} \\ \hline
A state $\psi$ is reached and is \textbf{possibly} preceded \textbf{at some time} by a state $\phi$ & $EF~ (\phi \wedge EF ~(\psi))$ & $\mu X . ((\phi \wedge \mu Y . (\psi \vee \lozenge Y)) \vee \lozenge X)$ \\ 
A state $\psi$ is reached and is \textbf{possibly} preceded \textbf{all the time} by a state $\phi$ & $E~ (\phi ~U~ \psi)$ & $\mu X . (\psi \vee (\phi \wedge \lozenge X))$ \\ 
A state $\psi$ is reached and is \textbf{necessarily} preceded \textbf{at some time} by a state $\phi$ & $EF~ (\psi) ~\wedge$ & $\mu X . (\psi \vee \lozenge X) ~\wedge$ \\
& $\neg E~ (\neg \phi ~U~ \psi)$ & $\neg \mu Y . (\psi \vee (\neg \phi \wedge \lozenge Y))$\\
A state $\psi$ is reached and is \textbf{necessarily} preceded \textbf{all the time} by a state $\phi$ & $EF~ (\psi) ~\wedge \neg E (T ~U $ & $\mu X . (\psi \vee \lozenge X) ~\wedge \neg \mu Y . (\neg \phi ~\wedge $ \\
& $(\neg \phi \wedge E (T ~U~ \psi)))$ & 
$
\mu Z . (\psi \vee (T \wedge \lozenge Z))
\vee (T \wedge \lozenge Y))$ \\ \hline
\hline

\multicolumn{3}{|l|}{\textbf{Invariance pattern}} \\ \hline
A state $\phi$ \textbf{can} persist indefinitely & $EG~ (\phi)$ & $\nu X . (\phi \wedge \lozenge X)$ \\
A state $\phi$ \textbf{must} persist indefinitely & $AG~ (\phi)$ & $\nu X . (\phi \wedge \square X)$ \\ \hline 

\end{tabular} \\

\caption{Rules for the translation of the patterns into CTL and $\mu$-calculus. For each of the four patterns, the translation of all variants is shown.
We use the version of $\mu$-calculus presented in~\cite{Kupferman00}, which is interpreted on classical Kripke structures. The symbol $T$ stands for True.}

\label{translation:ctl:mu}
}
\end{table}
\end{landscape}

\section{Carbon Starvation Response in \textit{E. coli}}
\label{sec:bioapplic}

\subsection{Model of carbon starvation response}

To test the applicability of the temporal logic patterns, we have used
our approach for the analysis of a model of the carbon starvation
response in the bacterium \textit{E. coli}. In the absence of
essential carbon sources in its growth environment, an
\textit{E. coli} population abandons exponential growth and enters a
non-growth state called stationary phase.
This growth-phase transition
is accompanied by numerous physiological changes in the bacteria \cite{Huisman}, and controlled on the molecular
level by a complex genetic regulatory network integrating various
environmental signals. 

The molecular basis of the adaptation of the growth of
\textit{E. coli} to the nutritional conditions has been the focus of
extensive studies for decades \cite{gutierrez_r_os:2007,Hengge-Aronis}. However,
notwithstanding the enormous amount of information accumulated on the
genes, proteins, and other molecules known to be involved in the
stress adaptation process, it is currently not understood how the
response of the cell emerges from the network of molecular
interactions. Moreover, with some exceptions \cite{bettenbrock},
numerical values for the kinetic parameters
and the molecular concentrations are absent, which makes it difficult
to apply traditional methods for the dynamical modeling of genetic
regulatory networks.

These circumstances have motivated the development of a
qualitative model of the carbon starvation response network using a class of
{\em piecewise-linear (PL) differential equations}. The PL models,
originally introduced by \cite{Kauffman}, provide a coarse-grained
picture of the dynamics of genetic regulatory networks.
They associate a protein concentration variable to each of the
genes in the network, and capture the switch-like character of gene
regulation by means of step functions that change their value at a
threshold concentration of the proteins. The advantage of using
PL models is that the qualitative dynamics of the high-dimensional systems are relatively
simple to analyze, using inequality
constraints on the parameters rather than exact numerical
values \cite{HdJ2427,paper-method2}. This makes the PL models a valuable tool
for the analysis of the carbon starvation network.

In previous work we developed a PL model that we extend here by the general stress response factor RpoS and related regulators (\cite{BioSystems};
Ropers \textit{et~al}., in preparation). The dynamics of this system are described by
nine coupled PL differential equations, and fifty inequality
constraints on the parameter values.

\subsection{Qualitative simulation of carbon starvation response}

The mathematical properties of the class of PL models used for
modeling the stress response network have been well-studied
\cite{Kauffman}. We have previously shown how discrete abstractions can be used to
convert the continuous dynamics of the PL systems into a FSTS
\cite{paper-method2}. The states $S$ of the FSTS correspond to
hyperrectangular regions in the concentration space, while the
transitions $T$ arise from trajectories that enter one region from
another. The atomic propositions $AP$ describe, among other things,
the concentration bounds of the regions and the trend of the variables
inside a region (increasing, decreasing, or steady). The generation of
the FSTS from the PL model has been implemented in the computer tool
Genetic Network Analyzer (GNA) \cite{HdJ2427}. GNA is able to export
the FSTS to standard model checkers like NuSMV \cite{nusmv} and CADP
\cite{Garavel-Lang-Mateescu-Serwe-07}.

The application of this approach to the model of the \textit{E. coli}
carbon starvation network generates a huge FSTS. The entire state set
consists of approximately $\mathcal{O}(10^{10})$ states, 
while the subset of states that is
most relevant for our purpose, \textit{i.e.} the states that are
reachable from an initial state corresponding to a particular growth state of the bacteria, still consists of $\mathcal{O}(10^3)$ states. 
It is clear that FSTSs of this size cannot be
analyzed by visual inspection, and that formal verification techniques
are needed.

In the next section we show how the
patterns defined in Sec.~\ref{sec:patterns,descr} can speed up the
querying of these FSTSs, by simplifying the formulation of relevant properties to be tested.
We are particularly interested in the question how the extension of the model with RpoS affects the predicted dynamics of the system.  The instances of the patterns
were translated into the temporal logic CTL following the translation
rules of Table~\ref{translation:ctl:mu}, and then verified using the
model-checker NuSMV.

\thispagestyle{empty}
\begin{sidewaysfigure}
\resizebox{0.97\textwidth}{!}{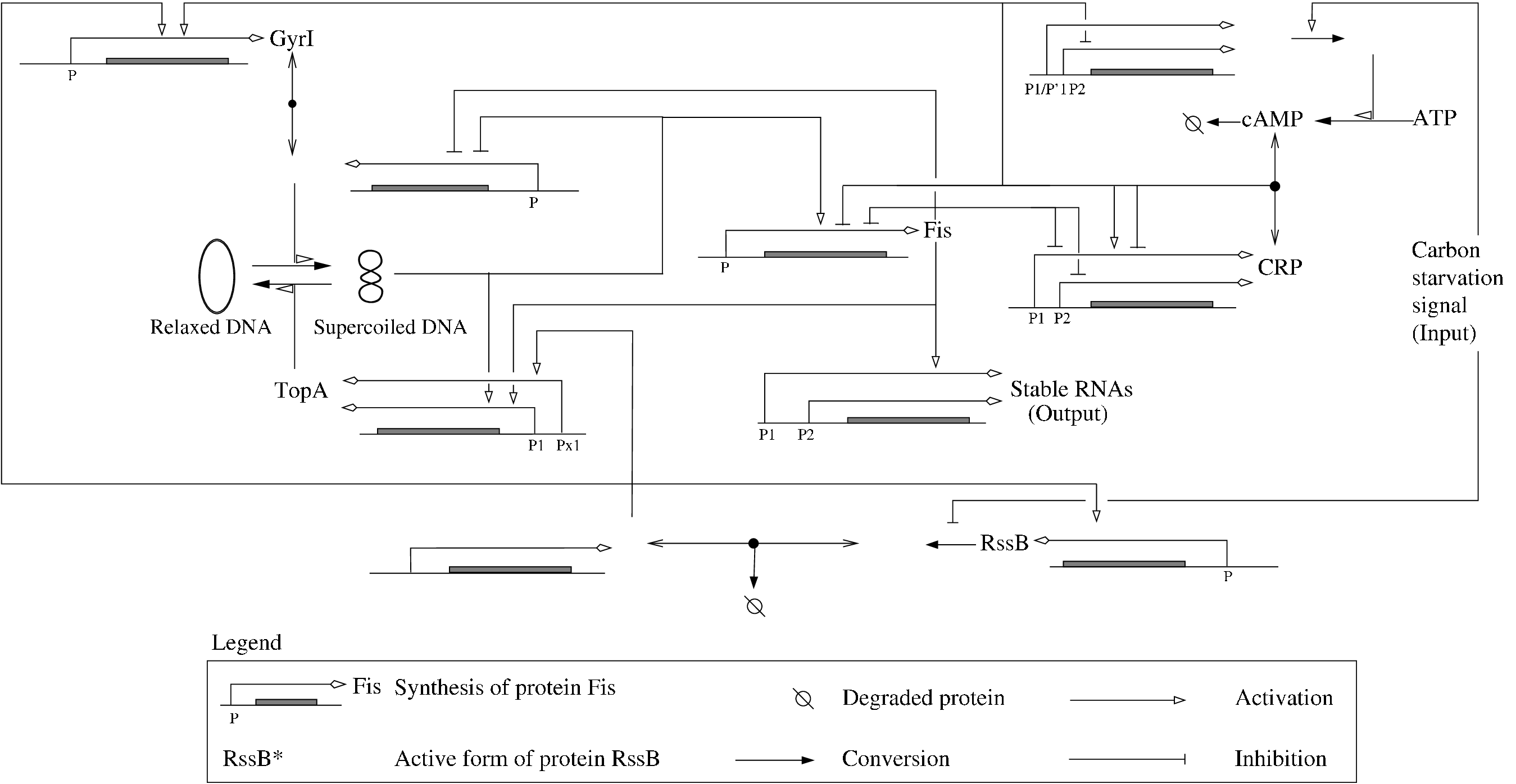}
\hfill\mbox{(a)} \\

$\dot{x}_{gyrAB} = \kappa_{gyrAB}\ (1 - s^{+}(x_{gyrAB},\theta_{gyrAB}^{2})\ s^{-}(x_{gyrI},\theta_{gyrI}^{1})\ s^{-}(x_{topA},\theta_{topA}^{1}))\
s^{-}(x_{fis},\theta_{fis}^4)-\gamma_{gyrAB}\ x_{gyrAB}$ \\
$0< \theta_{gyrAB}^{1} < \theta_{gyrAB}^{2} < \kappa_{gyrAB}/\gamma_{gyrAB} < \mathit{max}_{gyrAB}$
\hfill\mbox{(b)}

\caption{(a) Network of key genes, proteins and regulatory
interactions involved in the carbon starvation response network in
\textit{E. coli}. (b) PL differential equation and parameter
inequality constraints for the gyrase GyrAB. The variable $x_{gyrAB}$
denotes the concentration of GyrAB. The protein is produced at a rate
$\kappa_{gyrAB}$ if the DNA supercoiling level is not high, that is,
if the concentration of GyrAB itself is below the threshold
$\theta_{gyrAB}^2$, and the concentrations of the topoisomerase TopA
and the gyrase inhibitor GyrI are above the thresholds
$\theta_{topA}^1$ and $\theta_{gyrI}^1$, respectively. The regulatory logic of \textit{gyrAB} expression is modeled by means of step functions. For instance, $s^{+}(x_{gyrAB},\theta_{gyrAB}^{2})$ evaluates to 1, if $x_{gyrAB} >\theta_{gyrAB}^{2}$ (and to 0 otherwise). The protein is
degraded at a rate proportional to its own concentration,
$\gamma_{gyrAB}\ x_{gyrAB}$. The constraint $\theta_{gyrAB}^{2} <
\kappa_{gyrAB}/\gamma_{gyrAB} < \mathit{max}_{gyrAB}$ express that the
derepression of the \textit{gyrAB} promoter allows the concentration
of GyrAB to reach a high level, above the threshold
$\theta_{gyrAB}^{2}$.
\label{fig:network}}
\end{sidewaysfigure}

\section{Analysis of Carbon Starvation Response Model using query patterns}
\label{sec:analysis}

\subsection{Mutual inhibition of Fis and CRP}
The proteins Fis and CRP mutually inhibit their expression (Fig.~\ref{fig:network}). The regulatory protein CRP is the target of a signal-transduction pathway, which activates the adenylate cyclase Cya in case of carbon starvation. In turn, the latter synthesizes a small molecule, cAMP, which binds to CRP. This active form of CRP is able to regulate the expression of a large number of genes. In particular, CRP$\cdot$cAMP binds to the promoter region of the gene \textit{fis}, thereby preventing synthesis of new Fis proteins. Fis is an important regulator of genes involved in the cellular metabolism but it also inhibits \textit{crp} expression, by binding to multiple sites in the two promoter regions of the gene, P1 and P2. The regulatory interactions between genes \textit{fis} and \textit{crp} form a positive feedback loop, a motif often found in the genetic regulatory networks. When present in isolation, this kind of motif has been shown to lead to bistability \cite{Gardner-toggleswitch}.

The question can be asked whether the motif is also functional in the context of the carbon starvation response network. For instance, the expression of \textit{fis} is not only controlled by CRP$\cdot$cAMP, but also by the DNA supercoiling level and Fis itself. To check whether the bistability property is preserved in the larger network, we used two instances of the {\em occurrence/exclusion} pattern to express that it is impossible that proteins Fis and CRP be simultaneously present at high and at low concentration in the cells (Table~\ref{translation:biopatterns}). This property was shown to be true by the NuSMV model checker. 
We conclude that the positive feedback loop involving \textit{fis} and \textit{crp} is functional.

\subsection{Damped oscillations after carbon upshift}

The carbon starvation response network also contains a negative feedback loop, involving the genes \textit{gyrAB}, \textit{topA}, and \textit{fis}
(Fig.~\ref{fig:network}). GyrAB is a gyrase protein which supercoils the DNA structure, whereas the topoisomerase TopA relaxes it. An increase of the DNA supercoiling level stimulates expression of Fis, which in turn decreases the supercoiling level, by stimulating \textit{topA} expression and inhibiting \textit{gyrAB} expression.
The resulting negative feedback loop was predicted to give rise to (damped) oscillations of Fis and GyrAB concentrations after a carbon upshift \cite{BioSystems}.

In the present version of our model, additional interactions contribute to controlling the DNA supercoiling level. Hence, the gyrase inhibitor GyrI represses the activity of GyrAB by forming a complex with the protein.
The expression of \textit{gyrI} is notably stimulated by RpoS \cite{oh}. We formulated a {\em consequence} pattern to verify whether this affects the functioning of the negative feedback loop. In particular, we checked whether the carbon upshift is still a necessary condition for the occurrence of damped oscillations, as it was in the previous model (Table~\ref{translation:biopatterns}).
In the pattern we made use of the state descriptor \textit{isOscillatoryState}, which labels states as belonging to a (terminal) cycle in the FSTS. The occurrence of an oscillatory state could alternatively be expressed using temporal logic formulas (Sec.~\ref{sec:discussion}). The model-checker returned true for the query, meaning that the damped oscillations still occur following a carbon upshift.

\subsection{Control of entry into stationary phase by RpoS}

RpoS is a stress sigma factor that allows cells to adapt to and survive under harmful conditions by entering stationary phase \cite{Hengge-Aronis}.
Due to its key role, the concentration of RpoS is tightly regulated, at the transcriptional, translational, and post-translational levels.
The stability of the protein is mainly controlled in our conditions: while cells grow on a carbon source, RpoS is actively degraded through the protein RssB, which binds to RpoS and targets the factor to an intracellular protease.
However, the depletion of the carbon source inactivates RssB, thus allowing RpoS to accumulate at a high concentration.

Given the importance of RpoS for cell survival, one may ask whether the entry into stationary phase is always preceded by the accumulation of RpoS in the cell. We formulated this question using a \textit{sequence} pattern, where the stationary phase is represented by a low level of stable RNAs \textit{rrn} (Table~\ref{translation:biopatterns}). The latter indicator is motivated by the fact that stationary-phase cells do not need high levels of these RNAs, which are necessary for the high translational activity of the exponential phase. The property is true, which indicates that the entry into stationary phase cannot occur before RpoS has accumulated.

\subsection{Expression of \textit{topA} during growth-phase transitions}

Our previous model was incapable of accounting for the control of DNA supercoiling during growth-phase transitions. In particular, TopA was predicted to be never expressed, which is consistent with published data \cite{qi}. The extension of the model with RpoS makes it possible to refine the description of the control of the DNA supercoiling level. On the one hand, GyrAB activity is regulated by GyrI, as mentioned previously, and on the other hand, the \textit{topA} promoter is activated by RpoS.

In order to know whether \textit{topA} is expressed in response to the carbon source availability, we used an {\em invariance} pattern to check if the absence of \textit{topA} expression persists indefinitely (Table~\ref{translation:biopatterns}). The corresponding temporal logic formula is false and the diagnostic of the model-checker shows that expression of \textit{topA} is stimulated at the entry into stationary phase, most likely under the influence of RpoS. Indeed, following carbon starvation, the protein RssB is inactivated, which leads to the accumulation of RpoS at high levels. RpoS in turn stimulates the expression of \textit{topA}.

\begin{landscape}
\begin{table}
\begin{tabular}{|l|}
\hline
\textbf{Occurrence/exclusion pattern}: Mutual inhibition of Fis and CRP \\ \hline
$\mid$ It $\mid$ is not possible $\mid$ for a state $\mid$ $ x_{crp} \geq \frac{k^1_{crp}+k^2_{crp}+k^3_{crp}}{\gamma_{crp}} \wedge x_{fis} \geq \theta^4_{fis}$ $\mid$ to occur $\mid$ \\ $\mid$ It $\mid$ is not possible $\mid$ for a state $\mid$ $ x_{crp} \leq \frac{k^1_{crp}}{\gamma_{crp}} \wedge x_{fis} \leq \theta^1_{fis} $ $\mid$ to occur $\mid$ \\ \hline
CTL: $\neg EF ( x_{crp} \geq \frac{k^1_{crp}+k^2_{crp}+k^3_{crp}}{\gamma_{crp}} \wedge x_{fis} \geq \theta^4_{fis} ) $ 
$\wedge \neg EF ( x_{crp} \leq \frac{k^1_{crp}}{\gamma_{crp}} \wedge x_{fis} \leq \theta^1_{fis} )$ \\
$\mu$-calculus: $\neg \mu X . ((x_{crp} \geq \frac{k^1_{crp}+k^2_{crp}+k^3_{crp}}{\gamma_{crp}} \wedge x_{fis} \geq \theta^4_{fis}) \vee \lozenge X)$
$\wedge \neg \mu X . ((x_{crp} \leq \frac{k^1_{crp}}{\gamma_{crp}} \wedge x_{fis} \leq \theta^1_{fis}) \vee \lozenge X)$ \\
\hline
\hline
\textbf{Consequence pattern}: Damped oscillations after nutrient upshift \\ \hline
$\mid$ If a state $\mid$ $x_{signal} < \theta_{signal}$ $\mid$ occurs, then it is $\mid$ necessarily $\mid$ followed by a state $\mid$ $isOscillatoryState$ $\mid$ \\ \hline
CTL: $AG~ ((x_{signal} < \theta_{signal}) \Rightarrow AF~ (isOscillatoryState))$ \\
$\mu$-calculus: $\nu X . (((x_{signal} < \theta_{signal}) \Rightarrow \mu Y . (isOscillatoryState \vee \square Y)) \wedge \square X)$ \\
\hline
\hline
\textbf{Sequence pattern}: Control of entry into stationary phase by RpoS \\ \hline
$\mid$ A state $\mid$ $x_{rpoS} \geq \theta^1_{rpoS}$ $\mid$ is reached and is $\mid$ necessarily $\mid$ preceded $\mid$ at some time $\mid$ by a state $\mid$ $x_{rrn} > \theta_{rrn}$ $\mid$ \\ \hline
CTL: $EF~ (x_{rpoS} \geq \theta^1_{rpoS}) \wedge \neg E~ (\neg (x_{rrn} > \theta_{rrn}) ~U~ (x_{rpoS} \geq \theta^1_{rpoS}))$ \\
$\mu$-calculus: $\mu X . ((x_{rpoS} \geq \theta^1_{rpoS}) \vee \lozenge X) ~\wedge \neg \mu Y . ((x_{rpoS} \geq \theta^1_{rpoS}) \vee (\neg (x_{rrn} > \theta_{rrn}) \wedge \lozenge Y))$ \\
\hline
\hline
\textbf{Invariance pattern}: Expression of \textit{topA} during growth-phase transitions \\ \hline
$\mid$ A state $\mid$ $x_{topA} < \theta^1_{topA}$ $\mid$ can $\mid$ persist indefinitely $\mid$ \\ \hline
CTL: $EG~ ( x_{topA} < \theta^1_{topA} )$ \\
$\mu$-calculus: $\nu X . ((x_{topA} < \theta^1_{topA}) \wedge \lozenge X)$ \\
\hline 
\end{tabular} \\

\caption{Translation of the instances of the patterns used in the analysis of the \textit{E. coli} carbon starvation response into CTL and $\mu$-calculus, following the translation rules in Table~\ref{translation:ctl:mu}.}
\label{translation:biopatterns}
\end{table}
\end{landscape}

\section{Discussion}
\label{sec:discussion}

Formal verification techniques are promising tools for upscaling the analysis of cellular interaction networks. The widespread adoption of model-checking approaches is restrained, however, by the difficulty for non-expert users to formulate appropriate questions in temporal logics. Inspired by work in the formal verification community (\cite{HdJ2605}, see also \cite{Pnueli}), the first contribution of the paper consists in the formulation of a set of patterns in the form of query templates in structured natural language. The patterns capture a large number of frequently-asked questions by modelers in systems biology, but they are not restricted to a particular type of network or a particular biological system. In addition, we have provided translations of the patterns to two different temporal logics, CTL and $\mu$-calculus. 

The second contribution of the paper is the use of these patterns for the analysis of the complex genetic regulatory network involved in the carbon starvation response in \textit{E. coli}. We have modeled this network by means of PL differential equations and simulated the qualitative dynamics of the system in response to carbon starvation and carbon upshift. Our model extends a previous model \cite{BioSystems} with additional global regulators, notably the sigma factor RpoS, to better account for the control of DNA supercoiling during the growth transitions of the bacteria. The patterns are instantiated to verify the effect of this addition to the predicted network dynamics.

The patterns proposed in this paper are globally consistent with those discussed in \cite{HdJ2605}, but there are differences due to the specific application domain for which our patterns were developed.
For instance, the notion of scope used by \cite{HdJ2605} is not commonly defined for all the patterns, but implicitly present through the use of specific variants for each pattern.
Also, we have not explicitly included patterns that can be obtained by the recursive application of other patterns, such as the {\em chain response} pattern defined in \cite{HdJ2605}.
While patterns have not been used for the querying of cellular interaction networks thus far, some papers list example questions. It is reassuring to observe that all questions in the list of \cite{HdJ2380} can be expressed by means of the patterns in Sec.~\ref{sec:patterns,descr}.

An obvious generalization of the patterns proposed in this paper, already briefly mentioned in Sec.~\ref{sec:patterns}, would be to allow state descriptors that are formulas in temporal logic. For instance, instead of using atomic propositions to label states belonging to a (terminal) cycle in the FSTS, which requires the preliminary detection of strongly connected components in the state transition graph, we could use temporal logic formulas \cite{HdJ2427}. The introduction of temporal logic formulas as state descriptors makes the patterns more general, but also potentially more complicated to formulate and dependent on a particular temporal logic. A compromise trading some expressive power for user-friendliness would be to restrict the possible temporal logic formulas to simple forms and introduce these as predefined state descriptors (Sec.~\ref{sec:patterns,descr}). This is consistent with the main idea underlying the use of patterns, namely that they cannot be expected to cover all possible queries, but rather should allow users to formulate their frequent questions without worrying about the translation to temporal logic.

\newpage
\bibliographystyle{plain}
\bibliography{rapport-inria_patterns}

\newpage
\tableofcontents

\end{document}